\documentclass[%
 reprint,
 amsmath,amssymb,
 aps, superscriptaddress
]{revtex4-2}
\usepackage{ytableau}
\usepackage[hyperfootnotes=false]{hyperref}
\hypersetup{
 colorlinks=true,
 citecolor=blue,
 linkcolor=red,
 urlcolor=blue}

\usepackage{graphicx}
\usepackage{dcolumn}
\usepackage{bm}

\begin{document}

\preprint{APS/123-QED}

\title{Exotic SU(3) Flavor Structures in Fully Light Tetraquark Systems}
\thanks{ankushsharma2540.as@gmail.com}%

\author{Ankush Sharma}
\affiliation{Department of Physics, Chandigarh Engineering College (CEC), CGC Landran, Mohali, Punjab, India}
\author{Alka Upadhyay}
 \affiliation{Department of Physics and Material Science, Thapar Institute of Engineering and Technology, Patiala, Punjab, India}

\date{\today}

\begin{abstract}
The study of fully light tetraquark states composed solely of the light quarks \(u, d,\) and \(s\) provides an essential framework to understand the underlying dynamics of low-energy Quantum Chromodynamics (QCD). Within the framework of SU(3)\(_f\) flavor symmetry, these states are classified into different multiplets, giving rise to a rich spectrum of non-strange, singly strange, doubly strange, and hidden-strangeness configurations. In this work, we perform a comprehensive spectroscopic analysis of fully light tetraquarks using the extended Gursey–Radicati mass formula. The mass spectra are evaluated by incorporating spin-, isospin-, and hypercharge-dependent Casimir operators, with parameters extracted from the known meson and baryon sectors. The resulting predictions show clear mass hierarchies among the multiplets and are consistent with experimentally observed light scalar mesons such as \(f_0(500)\), \(a_0(980)\), and \(f_0(980)\). Furthermore, spin-parity assignments, mass splitting patterns, and possible mixing between octet and singlet configurations are discussed in detail. This analysis not only strengthens the theoretical basis for interpreting light scalar mesons as tetraquark candidates but also establishes the GR mass formula as an efficient phenomenological tool for exploring multiquark spectroscopy within SU(3)\(_f\) symmetry.
\end{abstract}


\maketitle



\section{Introduction}
The conventional quark model proposed independently by Gell-Mann and Zweig in 1964 successfully classified hadrons into mesons (\(q\bar{q}\)) and baryons (\(qqq\)) based on the SU(3) flavor symmetry \cite{GellMann1964,Zweig1964}. However, Quantum Chromodynamics, allows for more complex color-singlet configurations beyond these conventional structures, including tetraquarks (\(qq\bar{q}\bar{q}\)), pentaquarks (\(qqqq\bar{q}\)), and hybrid states containing valence gluons. These unconventional configurations, generally referred to as exotic hadrons, have attracted considerable theoretical and experimental attention due to recent discoveries of states that cannot be accommodated within the traditional quark model spectrum \cite{Chen2016,Esposito2017,Lebed2017,Guo2018}.

Among these, tetraquarks hold a particularly significant position in the ongoing exploration of multiquark dynamics. The earliest systematic investigation of tetraquarks was carried out by Jaffe using the MIT bag model, where he proposed the light scalar mesons below 1~GeV as compact \(qq\bar{q}\bar{q}\) states \cite{Jaffe1977}. Since that pioneering work, theoretical studies have developed along multiple directions, employing a range of frameworks such as QCD sum rules, lattice QCD, potential models, effective Hamiltonian approaches, and phenomenological relations like the Gursey–Radicati mass formula and its modern extensions \cite{Close2002,Maiani2004,Ali2017}. These frameworks collectively aim to explain the mass hierarchy, decay mechanisms, and internal structure of tetraquarks while differentiating compact diquark–antidiquark configurations from meson–meson molecular-like states.

While experimental evidence for heavy and hidden-heavy tetraquarks has grown steadily in recent years—particularly through the LHCb collaboration’s identification of charmonium-like and bottomonium-like resonances \cite{LHCb2020, Aaij2021} the light tetraquark sector remains theoretically challenging and experimentally elusive. Fully light tetraquarks, composed exclusively of \(u\), \(d\), and \(s\) quarks, provide a unique laboratory for exploring the strongly coupled, non-perturbative regime of QCD. In contrast to heavy-flavor systems, where non-relativistic approximations are reasonably valid, the light quark sector exhibits strong chiral symmetry breaking, flavor mixing, and dynamical coupling to meson–meson continua, rendering theoretical predictions considerably more intricate \cite{Oller2000,Black1999,Tornqvist1995}.

The study of fully light tetraquarks is not merely an extension of heavy-quark spectroscopy but rather an essential probe of the mechanisms governing color confinement and chiral dynamics. Several low-lying scalar mesons, such as \(f_0(500)\), \(f_0(980)\), \(a_0(980)\), and \(K_0^*(700)\), have long been hypothesized as candidates for tetraquark or dynamically generated molecular states \cite{Jaffe1977,Weinstein1990,Pelaez2016}. Disentangling these possibilities requires a precise understanding of their spectroscopic patterns, decay properties, and quantum number assignments. In this context, identifying the correct internal configuration—whether compact diquark–antidiquark or meson–meson correlated—remains one of the most intriguing and unresolved questions in light hadron spectroscopy.

To achieve this, theoretical models based on effective quark–mass relations have proven useful in providing quantitative insight. The extended Gursey–Radicati mass formula, originally formulated to describe baryon mass splittings \cite{Gursey1958,Radicati1965}, has been successfully generalized to exotic hadrons by including explicit spin, isospin, and hypercharge-dependent terms \cite{Anisovich2005,Chen2017}. This approach enables the systematic evaluation of tetraquark mass spectra and internal symmetry-breaking effects arising from SU(3)\(_f\) flavor symmetry violations. Within this framework, the light tetraquark configurations emerging from the flavor combination \(3 \otimes 3 \otimes \bar{3} \otimes \bar{3}\) yield exotic flavor multiplets, which form the foundation for spectroscopic analysis. In Ref. \cite{REF1}, fully light tetraquarks are classified in allowed multiplets using SU(3) representation.

In the present work, we perform a comprehensive investigation of fully light tetraquark systems constructed from combinations of light quarks only. The analysis employs the extended Gursey–Radicati mass relation to calculate their mass spectra, spin–parity (\(J^P\)) assignments, and flavor configurations. In our prevoius works, the application of the Gursey-Radicati mass formula to exotic systems has been quite successful in predicting their mass spectra \cite{Sharma:2023lij, Sharma:2023wnd, Sharma:2024ern, Sharma:2024pfi, Sharma:2025nrw, Sharma:2025A, Sharma:2025grd}. Furthermore, we explore their decay dynamics through possible OZI-allowed two-body decay channels. The proximity of tetraquark masses to relevant meson–meson thresholds is analyzed to understand the extent of configuration mixing between compact and molecular-like components. The resulting mass distributions and decay characteristics provide valuable theoretical guidance for forthcoming experiments at LHCb, BESIII, Belle II, which are expected to yield precision data on light meson resonances.

The study of fully light tetraquarks thus represents a crucial step toward deciphering the non-perturbative aspects of QCD. By systematically examining their structure, flavor organization, and decay patterns within a phenomenological yet symmetry-driven framework, this work aims to advance the theoretical understanding of exotic hadronic matter and bridge the gap between effective model predictions and experimental observables in the low-energy hadron sector. This manuscript is organized as follows. Section II presents the theoretical framework for the spectroscopy of fully light tetraquarks, including the classification scheme based on SU(3) flavor representations, construction of the total wavefunction involving flavor, spin, color, and spatial degrees of freedom, followed by the extension of the Gursey-Radicati mass formula, which accounts for the mass spectra of fully-light tetraquarks. Section III is devoted to the analysis of the mass spectra of fully light tetraquark systems. In Section IV, the possible production mechanisms and strong decay channels of the predicted states are discussed. Finally, Section V summarizes the main results and conclusions of the present work.

\section{Theoretical Framework for Fully-Light Tetraquarks}
In this section, we present the theoretical framework employed for the spectroscopic study of fully light tetraquark states composed of light quarks $(u,d,s)$. The non-perturbative nature of Quantum Chromodynamics in the light-quark sector makes the investigation of such multiquark systems particularly important for understanding exotic hadron spectroscopy. The tetraquark states are classified by constructing the total flavor, spin, and color wavefunctions under the SU(3) flavor symmetry, subject to the Pauli exclusion principle and overall color singlet constraint. This classification allows the identification of the relevant multiplet structures and their corresponding quantum numbers.

The mass spectra are evaluated using an extension of the Gursey-Radicati mass formula that incorporates spin-dependent interactions, flavor-symmetry-breaking effects, and group-theoretical contributions via Casimir operators. This formalism provides a systematic description of mass splittings among states with different spin and flavor configurations. To further investigate their experimental signatures, the strong-decay properties of the predicted tetraquark states are analyzed via their dominant two-meson fall-apart decay channels. The combined study of mass spectra and decay behavior provides important insight into the stability and possible experimental observation of fully light tetraquark states.

\subsection{Classification Scheme}
The investigation of fully light tetraquark systems of the form $qq\bar{q}\bar{q}$ is important for understanding the non-perturbative dynamics of Quantum Chromodynamics. Unlike conventional mesons $(q\bar{q})$ and baryons $(qqq)$, tetraquarks represent exotic multiquark configurations whose internal structure is governed by a delicate interplay of flavor symmetry, color confinement, and spin-dependent interactions. In the light quark sector, these effects become particularly significant due to strong relativistic corrections and enhanced flavor mixing. Therefore, a systematic classification of fully light tetraquarks is essential for identifying possible exotic resonances beyond the conventional quark model.

The total wavefunction of a tetraquark system must satisfy the Pauli exclusion principle for identical fermions. Since the system contains two quarks and two antiquarks, the symmetry constraints must be imposed separately on the quark pair $(qq)$ and the antiquark pair $(\bar{q}\bar{q})$. The complete wavefunction is written as
\begin{equation}
\Psi_{\text{total}}
=
\Psi_{\text{space}}
\otimes
\Psi_{\text{flavor}}
\otimes
\Psi_{\text{spin}}
\otimes
\Psi_{\text{color}},
\end{equation}
where each component carries specific symmetry properties. For the ground-state configuration, we consider zero orbital angular momentum $(l=0)$, which corresponds to the lowest-energy state and implies positive parity. In this case, the spatial wavefunction is symmetric,
\begin{equation}
\Psi_{\text{space}}
=
\Psi_{\text{space}}^{(S)}.
\end{equation}
As a consequence, the combined flavor-spin-color part must ensure the required anti-symmetry under the exchange of identical quarks.

The flavor structure of fully light tetraquarks is described within the SU(3) flavor symmetry generated by the light quarks $u$, $d$, and $s$. Each quark transforms as the fundamental representation $3$, while each antiquark belongs to the conjugate representation $\bar{3}$. Therefore, the flavor decomposition of the tetraquark system is given by
\begin{equation}
(3 \otimes 3)
\otimes
(\bar{3} \otimes \bar{3})
=
(6 \oplus \bar{3})
\otimes
(\bar{6} \oplus 3).
\end{equation}

This decomposition generates several possible SU(3) multiplets such as singlet $(1)$, octet $(8)$, decuplet-like structures, and the 27-plet. Among these, the 27-plet is of particular interest because it corresponds to the highest symmetric representation arising from the direct product of the symmetric diquark sextet $(6)$ and the symmetric antidiquark antisextet $(\bar{6})$. It contains states with genuinely exotic flavor quantum numbers that cannot be accommodated within the conventional $q\bar{q}$ meson picture. This makes the 27-plet an ideal candidate for studying exotic spectroscopy and for identifying possible experimentally observable nonconventional resonances.

The SU(3) 27-plet specifically emerges from
\begin{equation}
6 \otimes \bar{6}
=
1 \oplus 8 \oplus 27.
\end{equation}
This indicates that both the quark pair and the antiquark pair must belong to symmetric flavor representations,
\begin{equation}
\Psi_{\text{flavor}}^{qq}
=
\Psi_{\text{flavor}}^{(6)},
\qquad
\Psi_{\text{flavor}}^{\bar{q}\bar{q}}
=
\Psi_{\text{flavor}}^{(\bar{6})}.
\end{equation}
The flavor symmetry of the 27-plet is therefore strongly constrained and directly influences the allowed spin and color configurations through the Pauli principle.

The color degree of freedom is equally important because physical hadrons must be color singlets due to confinement. For the quark pair, the color decomposition follows
\begin{equation}
3 \otimes 3
=
\bar{3}_{A}
\oplus
6_{S},
\end{equation}
while for the antiquark pair,
\begin{equation}
\bar{3} \otimes \bar{3}
=
3_{A}
\oplus
\bar{6}_{S}.
\end{equation}

Although both symmetric and antisymmetric color structures are mathematically allowed, the antisymmetric diquark configuration $\bar{3}_{A}$ is generally favored dynamically. This preference arises from the attractive nature of one-gluon exchange interactions, which lower the energy of antisymmetric color states compared to symmetric sextet configurations. Consequently, the diquark--antidiquark picture is constructed using
\begin{equation}
\bar{3}
\otimes
3
=
1
\oplus
8,
\end{equation}
and the physical tetraquark state is identified with the singlet component. Therefore, we take
\begin{equation}
\Psi_{\text{color}}^{qq}
=
\bar{3}_{A},
\qquad
\Psi_{\text{color}}^{\bar{q}\bar{q}}
=
3_{A}.
\end{equation}

Since the spatial wavefunction is symmetric and the flavor wavefunction for the 27-plet is also symmetric, the Pauli exclusion principle requires the combined spin-color part of the quark pair to be antisymmetric,
\begin{equation}
\Psi_{\text{spin}}^{qq}
\otimes
\Psi_{\text{color}}^{qq}
=
A,
\end{equation}
and similarly for the antiquark pair. Because the chosen color configuration is antisymmetric, the spin wavefunction must be symmetric. This immediately fixes the diquark and antidiquark spins to
\begin{equation}
S_{qq}
=
1,
\qquad
S_{\bar{q}\bar{q}}
=
1.
\end{equation}

The total spin of the tetraquark is obtained by coupling these two spin-1 subsystems,
\begin{equation}
1 \otimes 1
=
0
\oplus
1
\oplus
2,
\end{equation}
which leads to three possible total spin states,
\begin{equation}
S
=
0,\ 1,\ 2.
\end{equation}
These correspond to scalar, axial-vector, and tensor tetraquark configurations, respectively.

Among the allowed spin configurations, the tensor state with $S=2$ is of particular interest for the present study. In the diquark--antidiquark picture, the coupling of two symmetric spin-1 constituents naturally gives rise to the higher-spin configuration, which can play an important role in the spectroscopy of exotic multiquark states. Such tensor states are especially relevant in identifying excited tetraquark resonances and understanding their spectroscopic patterns beyond the scalar ground state. Since both the diquark and antidiquark are taken with spin $S_{qq}=1$ and $S_{\bar{q}\bar{q}}=1$, their coupling allows the maximum total spin contribution corresponding to the tensor configuration. Therefore, for the fully light tetraquark belonging to the SU(3) 27-plet, we focus on the spin-2 assignment with quantum numbers $J^{P} = 2^{+}$.

Finally, the complete wavefunction of the fully light tetraquark in the SU(3) 27-plet can be expressed as
\begin{equation}
\Psi_{\text{total}}
=
\Psi_{\text{space}}^{(S)}
\otimes
\Psi_{\text{flavor}}^{(27)}
\otimes
\Psi_{\text{spin}}^{(S)}
\otimes
\Psi_{\text{color}}^{(A)}.
\end{equation}
This construction satisfies the Pauli exclusion principle, preserves overall color neutrality, and provides the spectroscopic foundation for calculating the mass spectra and strong decay properties of fully light tetraquark states.

\begin{figure*}
    \centering
\includegraphics[width=0.6\linewidth]{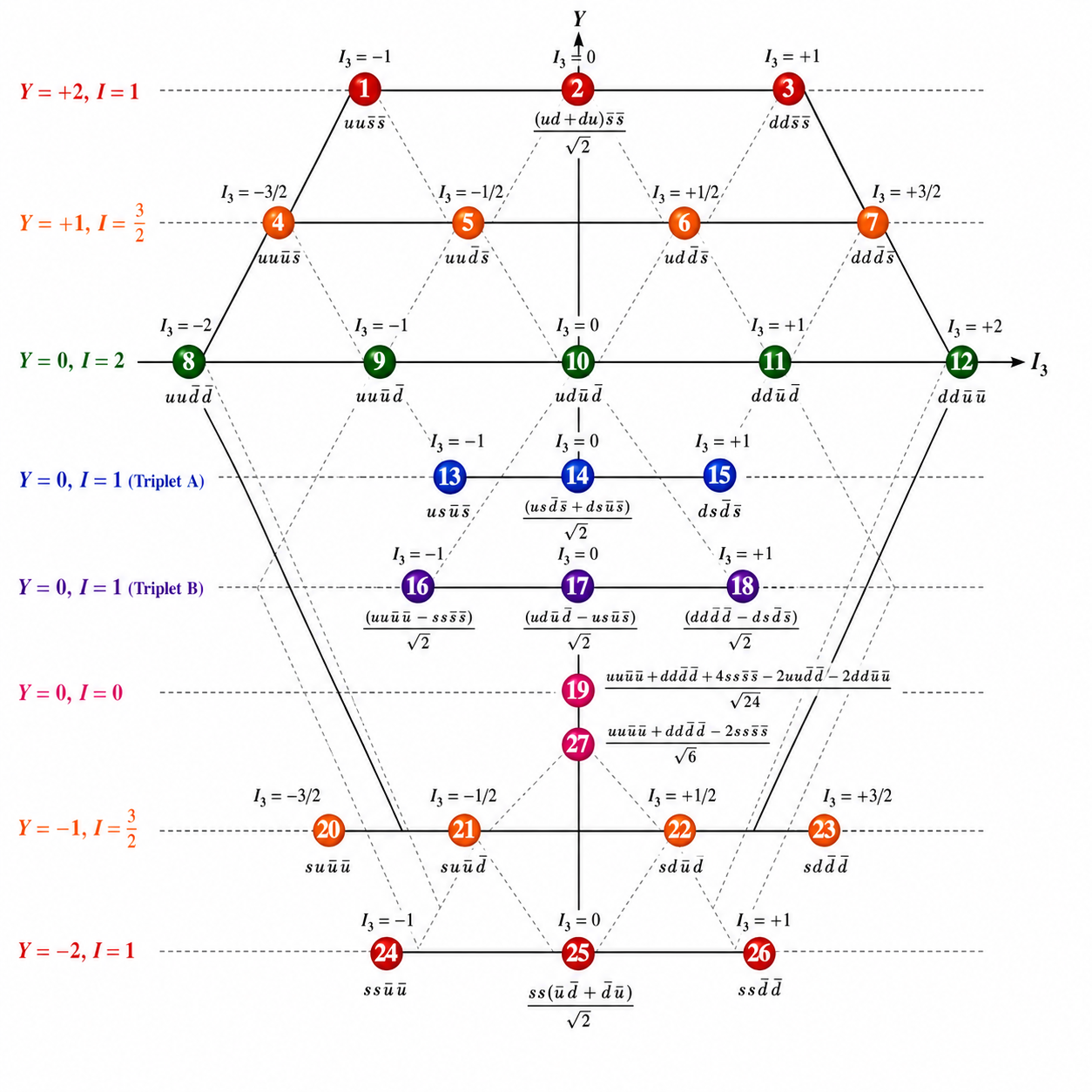}
    \caption{Weight diagram of the SU(3)$_F$ flavor $\mathbf{27}$-plet for fully light tetraquark states $(qq\bar{q}\bar{q})$, showing the distribution of states in the $(Y, I_3)$ plane with corresponding hypercharge $Y$, isospin $I$, and isospin projection $I_3$. The multiplet contains the exotic configurations with $(Y,I)=(+2,1)$, $(+1,\frac{3}{2})$, $(0,2)$, two independent $(0,1)$ triplets, the central $(0,0)$ states, $(-1,\frac{3}{2})$, and $(-2,1)$, completing the irreducible SU(3) flavor $\mathbf{27}$-plet structure.}
    \label{fig:placeholder}
\end{figure*}

\subsection{The Gursey-Radicati Mass Formula}
The Gursey-Radicati (GR) mass formula is one of the most successful phenomenological approaches for describing hadron mass splittings based on symmetry principles. It was originally proposed within the framework of broken $SU(6)$ spin-flavor symmetry to explain the observed baryon spectrum by incorporating the effects of spin, isospin, and hypercharge \cite{Gursey}. The central idea of the GR formalism is that hadron masses are not determined solely by constituent quark masses, but also receive important contributions from internal symmetry-breaking interactions, particularly hyperfine spin-dependent effects.

In its original form, the mass of a baryon is expressed as
\begin{equation}
M = M_{0} + aJ(J+1) + bY + c\left[T(T+1) - \frac{1}{4}Y^{2}\right],
\label{eq:GR_original}
\end{equation}
where $J$ denotes the total spin, $T$ is the isospin, and $Y$ represents the hypercharge of the baryon. The parameters $a$, $b$, and $c$ are phenomenological constants determined by fitting the experimentally observed baryon masses, while $M_{0}$ provides the overall mass scale.

Although Eq.~(\ref{eq:GR_original}) successfully describes conventional baryons, it becomes insufficient when applied to systems containing heavy quarks or exotic multiquark configurations. In such cases, additional flavor-dependent contributions must be included to account for the distinct dynamics of charm and bottom quarks. To improve the predictive power of the model, the GR mass formula was generalized by introducing the quadratic Casimir operators of the relevant symmetry groups \cite{Bijker}, leading to
\begin{align}
M_{B} = M_{0} &+ A\,S(S+1) - B\,C_{2}(SU(6))
+ D\,Y  \nonumber \\
&+ G\,C_{2}(SU(3))
+ E\left[I(I+1) - \frac{1}{4}Y^{2}\right],
\label{eq:GR_generalized}
\end{align}
where $S$ is the total spin, $I$ is the isospin, and $C_{2}(SU(3))$ is the eigenvalue of the quadratic Casimir operator corresponding to the flavor $SU_{f}(3)$ representation. This term plays an important role in distinguishing hadrons belonging to different flavor multiplets.

For hidden-charm and multiquark systems such as tetraquarks and pentaquarks, further modifications are required. In particular, the formula must explicitly include the contribution of heavy quarks and antiquarks. To study the mass splitting among different $SU(3)$ multiplets of hidden-charm pentaquarks, an extended form of the GR mass formula was introduced as \cite{Santo}
\begin{align}
M_{GR} = M_{0} &+ A\,S(S+1) + D\,Y + E\left[I(I+1) - \frac{1}{4}Y^{2}\right]
\nonumber \\
&+ G\,C_{2}(SU(3))
+ F\,N_{c},
\label{eq:GR_hidden_charm}
\end{align}
where $N_{c}$ denotes the total number of charm quarks and anticharm antiquarks present in the system. The parameter $F$ quantifies the effective heavy-quark mass contribution.

To extend the formalism further for systems containing both charm and bottom quarks, the counter term was generalized by separating the heavy-flavor contributions \cite{HOLMA}. The final working form of the mass formula becomes
\begin{align}
M_{GR} = \xi M_{0}
&+ A\,S(S+1)
+ D\,Y + E\left[I(I+1) - \frac{1}{4}Y^{2}\right]
\nonumber \\
&+ G\,C_{2}(SU(3))
+ \sum_{i=c,b} F_{i}N_{i},
\label{eq:mass_formula}
\end{align}
where $N_{i}$ represents the number of heavy quarks (or antiquarks) of flavor $i = c,b$, and $F_{i}$ gives the corresponding heavy-flavor contribution. The parameter $\xi$ is introduced as a normalization factor that improves the description of multiquark systems by effectively accounting for higher-order corrections and overall mass renormalization effects.

In Eq.~(\ref{eq:mass_formula}), the parameter $M_{0}$ sets the basic mass scale and is associated with the total quark content of the system. The quantities $S$, $I$, and $Y$ correspond to the total spin, isospin, and hypercharge, respectively, while the quadratic Casimir operator $C_{2}(SU(3))$ determines the flavor multiplet structure. The term involving $F_{i}N_{i}$ explicitly incorporates heavy-quark symmetry-breaking effects, which are essential for accurately describing charm- and bottom-rich exotic states.

\begin{table}
\centering
\caption{Parameters of the extended Gursey-Radicati mass formula with their corresponding uncertainties \cite{Santo}.}
\label{tab:GR_parameters}
\tabcolsep=0.5mm
\begin{tabular}{ccccccc}
\hline
 & $M_{0}$ & $A$ & $D$ & $E$ & $G$ & $F_{c}$ \\
\hline
Values (MeV) 
& 940.0 
& 23.0 
& -158.3 
& 32.0 
& 52.5 
& 1354.6  \\
\hline
Uncertainty (MeV) 
& 1.5 
& 1.2 
& 1.3 
& 1.3 
& 1.3 
& 18.2 
\\
\hline
\end{tabular}
\end{table}

\begin{table*}[htbp]
\centering
\caption{Mass spectra of fully light tetraquark states belonging to the SU(3) flavor $\mathbf{27}$-plet having symmetric spin-parity = $2^+$ listed along with respective quantum numbers using the extension of the Gursey-Radicati mass formula.}
\label{tab:27plet_mass}
\small
\renewcommand{\arraystretch}{2.5}
\setlength{\tabcolsep}{6pt}

\begin{tabular}{ccccc}
\hline
Quark Contents of 27-plet states & $I$ & $Y$ & $C_2\big(SU(3)\big)$ & Predicted Mass (MeV) \\
\hline

$uu\bar{s}\bar{s}$, $\dfrac{(ud+du)\bar{s}\bar{s}}{\sqrt{2}}$, $dd\bar{s}\bar{s}$ & 1 & +2 & 8 & 1840.00 $\pm$ 13.06\\

$uu\bar{u}\bar{s}$, $uu\bar{d}\bar{s}$, $ud\bar{d}\bar{s}$, $dd\bar{d}\bar{s}$ & $\tfrac{3}{2}$ & +1 & 8 & 2078.30 $\pm$ 13.58 \\

$uu\bar{d}\bar{d}$, $uu\bar{u}\bar{d}$, $ud\bar{u}\bar{d}$, $dd\bar{u}\bar{d}$, $dd\bar{u}\bar{u}$ & 2 & 0 & 8 & 2316.60 $\pm$ 14.93 \\

$us\bar{u}\bar{s}$,$\dfrac{us\bar{d}\bar{s}+ds\bar{u}\bar{s}}{\sqrt{2}}$, $ds\bar{d}\bar{s}$ & 1 & 0 & 8 &  2188.60 $\pm$ 13.00\\

$\dfrac{uu\bar{u}\bar{u}-ss\bar{s}\bar{s}}{\sqrt{2}}$, $\dfrac{ud\bar{u}\bar{d}-us\bar{u}\bar{s}}{\sqrt{2}}$, $\dfrac{dd\bar{d}\bar{d}-ds\bar{d}\bar{s}}{\sqrt{2}}$,  & 1 & 0 & 8 & 2188.60 $\pm$ 13.00 \\

$\dfrac{uu\bar{u}\bar{u}+dd\bar{d}\bar{d}+4ss\bar{s}\bar{s}-2uu\bar{d}\bar{d}-2dd\bar{u}\bar{u}}{\sqrt{24}}$ & 0 & 0 & 8 & 2124.60 $\pm$ 12.73 \\

$su\bar{u}\bar{u}$, $su\bar{u}\bar{d}$, $sd\bar{u}\bar{d}$, $sd\bar{d}\bar{d}$ & $\tfrac{3}{2}$ & -1 & 8 & 2394.90 $\pm$ 13.58 \\

$ss\bar{u}\bar{u}$, $\dfrac{ss(\bar{u}\bar{d}+\bar{d}\bar{u})}{\sqrt{2}}$, $ss\bar{d}\bar{d}$ & 1 & -2 & 8 & 2473.20 $\pm$ 13.06 \\

$\dfrac{uu\bar{u}\bar{u}+dd\bar{d}\bar{d}-2ss\bar{s}\bar{s}}{\sqrt{6}}$ & 0 & 0 & 8 & 2124.60 $\pm$ 12.73 \\

\hline
\end{tabular}
\end{table*}

The coefficients $A$, $D$, $E$, $G$, $F_{c}$, and $F_{b}$ are determined through a $\chi^{2}$ minimization procedure by fitting the experimentally established masses of ground-state baryons, charmed baryons, and non-strange baryons. This fitting ensures the optimal agreement between theoretical predictions and available experimental data. The resulting parameter values, together with their corresponding uncertainties, are listed in Table~\ref{tab:GR_parameters}. Using these fitted parameters, the extended GR mass formula can be applied to predict the mass spectra of hidden-charm tetraquarks, singly heavy tetraquarks, and pentaquark systems containing four and five heavy quarks.

\section{Analysis of the Mass Spectra}
The mass spectra of fully light tetraquark states belonging to the SU(3) flavor $\mathbf{27}$-plet with symmetric spin-parity assignment $J^{P}=2^{+}$ are presented in Table \ref{tab:27plet_mass}. The masses have been evaluated using the extension of the Gürsey--Radicati mass formula by incorporating the effects of flavor symmetry, isospin, and group-theoretical contributions through the quadratic Casimir operator $C_{2}(SU(3))$. Since all the states belong to the same SU(3) multiplet, the value of the Casimir operator remains fixed at $C_{2}(SU(3))=8$, while the observed mass splittings arise primarily due to variations in hypercharge and flavor composition.

The obtained results exhibit a systematic increase in mass with increasing strange quark content, which reflects the explicit breaking of SU(3) flavor symmetry caused by the heavier strange quark mass. The lightest states of the multiplet are found in the $Y=+2$ sector with isospin $I=1$, corresponding to the configurations $uu\bar{s}\bar{s}$, $(ud+du)\bar{s}\bar{s}/\sqrt{2}$, and $dd\bar{s}\bar{s}$, having a predicted mass of $1840.00 \pm 13.06~\mathrm{MeV}$. These states form the uppermost members of the flavor multiplet and possess the minimum hidden-strangeness contribution among the exotic configurations considered.

As the hypercharge decreases from $Y=+2$ to $Y=0$, the masses increase gradually. The states with $I=\frac{3}{2}$ and $Y=+1$ are predicted around $2078.30 \pm 13.58~\mathrm{MeV}$, whereas the maximally charged isotensor configurations with $I=2$ and $Y=0$ appear at a significantly larger mass of $2316.60 \pm 14.93~\mathrm{MeV}$. The comparatively higher masses of these isotensor states indicate stronger flavor excitation effects within the fully light tetraquark system.

Interestingly, the neutral configurations belonging to the $Y=0$ sector exhibit partial degeneracy. The states containing mixed flavor structures such as $us\bar{u}\bar{s}$ and $(us\bar{d}\bar{s}+ds\bar{u}\bar{s})/\sqrt{2}$ are predicted at $2188.60 \pm 13.00~\mathrm{MeV}$, while the flavor-singlet dominated combinations are obtained near $2124.60 \pm 12.73~\mathrm{MeV}$. This behavior suggests that the interplay between flavor mixing and symmetry-breaking terms plays an important role in determining the internal dynamics of the tetraquark configurations.

In the negative hypercharge region, the masses continue to increase systematically. The states with $Y=-1$ and $I=\frac{3}{2}$ are predicted around $2394.90 \pm 13.58~\mathrm{MeV}$, whereas the doubly strange configurations with $Y=-2$ and $I=1$ attain the highest masses in the multiplet, namely $2473.20 \pm 13.06~\mathrm{MeV}$. The large masses of these states can be attributed to the presence of multiple strange quarks, which enhances the contribution from flavor symmetry breaking terms in the mass formula.

Overall, the predicted spectra display a well-defined hierarchical structure governed primarily by hypercharge and strangeness content. The relatively narrow theoretical uncertainties obtained for all states indicate the stability and consistency of the adopted formalism. The present results provide useful theoretical guidance for future experimental searches for fully light exotic tetraquark states in the mass region between $1.8$ and $2.5~\mathrm{GeV}$.

\section{Production Mechanism and Strong Decay Modes}

The fully light tetraquark states belonging to the SU(3) flavor $\mathbf{27}$-plet with symmetric spin-parity assignment $J^{P}=2^{+}$ can predominantly be produced in high-energy hadronic interactions, radiative decays of heavy quarkonia, central exclusive production processes, and relativistic heavy-ion collisions where multiquark configurations are expected to be enhanced due to the large density of light quarks and antiquarks. In particular, proton--proton collisions at LHC energies, photoproduction experiments, and $J/\psi$ radiative decays provide favorable environments for the formation of exotic light tetraquark states through gluon-rich intermediate processes. Since the present states possess exotic flavor configurations and higher isospin values, their observation through conventional $q\bar q$ meson channels is expected to be suppressed, thereby making them important candidates for identifying genuine multiquark structures.

The dominant strong decay mechanism for the fully light tetraquark states is expected to proceed through the OZI-superallowed fall-apart process into two conventional light mesons. The decay channels are determined by conserving isospin, hypercharge, charge, angular momentum, and parity. Owing to the predicted mass range between approximately $1.8$ and $2.5~\mathrm{GeV}$, most states are kinematically allowed to decay into pairs of pseudoscalar and vector mesons.

The exotic states with hypercharge $Y=+2$ and isospin $I=1$, corresponding to the configurations
\[
uu\bar{s}\bar{s}, \qquad \frac{(ud+du)\bar{s}\bar{s}}{\sqrt{2}}, \qquad dd\bar{s}\bar{s},
\]
represent manifestly exotic tetraquark structures since such flavor combinations cannot be accommodated within the conventional $q\bar q$ meson picture. These states possess double antistrangeness and therefore constitute excellent candidates for identifying genuine multiquark configurations. Experimentally, these states may be produced in high-energy hadronic collisions through gluon-rich processes involving strange quark pair creation,
\[
gg \rightarrow s\bar s s\bar s \rightarrow qq\bar q \bar q,
\]
particularly in proton--proton collisions at LHC energies and in heavy-ion environments where abundant strange quark production is expected. In addition, kaon-induced reactions such as
\[
K^{+}p \rightarrow X^{++}+K^{-},
\]
and photoproduction processes involving strange mesons may also provide favorable production mechanisms for these exotic configurations.

The quark rearrangement mechanism suggests that these states predominantly decay through fall-apart processes into two kaonic mesons. The dominant strong decay channels are expected to be
\[
K K,\qquad K K^{*},\qquad K^{*}K^{*},
\]
where the vector-vector decay mode may become particularly significant due to the symmetric spin assignment $J^{P}=2^{+}$. The state $uu\bar{s}\bar{s}$ may decay into channels such as
\[
K^{+}K^{+},\qquad K^{+}K^{*+},\qquad K^{*+}K^{*+},
\]
while the neutral member
\[
\frac{(ud+du)\bar{s}\bar{s}}{\sqrt{2}},
\]
can decay into
\[
K^{+}K^{0},\qquad K^{+}K^{*0},\qquad K^{*+}K^{0},
\]
and related vector-meson combinations. Similarly, the $dd\bar{s}\bar{s}$ configuration mainly couples to
\[
K^{0}K^{0},\qquad K^{0}K^{*0},\qquad K^{*0}K^{*0}.
\]
Since these states carry exotic flavor quantum numbers, mixing with ordinary mesons is expected to be strongly suppressed, which may help in their experimental identification through clean kaon-pair invariant mass spectra.

The states with hypercharge $Y=+1$ and isospin $I=\frac{3}{2}$,
\[
uu\bar{u}\bar{s}, \qquad uu\bar{d}\bar{s}, \qquad ud\bar{d}\bar{s}, \qquad dd\bar{d}\bar{s},
\]
also possess exotic isospin assignments which cannot arise in ordinary mesonic systems. Their production may occur through associated strange meson production in hadronic collisions, particularly in processes involving strong quark rearrangement and gluon fragmentation. Reactions such as
\[
pp \rightarrow X + K + \pi,
\]
or
\[
\gamma p \rightarrow X + K,
\]
may provide suitable production environments for these states. Because these configurations contain a single strange antiquark, they can be copiously produced in strange-rich hadronic interactions.

The dominant strong decay channels of these states are expected to proceed through
\[
\pi K,\qquad \pi K^{*},\qquad \rho K,\qquad \rho K^{*},
\]
final states. The configuration $uu\bar{d}\bar{s}$, for instance, can decay strongly into
\[
\pi^{+}K^{+},\qquad \rho^{+}K^{+},\qquad \pi^{+}K^{*+},
\]
through simple quark recombination. Similarly, the state $ud\bar{d}\bar{s}$ may couple to
\[
\pi^{0}K^{+},\qquad \pi^{+}K^{0},\qquad \rho^{0}K^{+},
\]
whereas $dd\bar{d}\bar{s}$ predominantly decays into
\[
\pi^{-}K^{0},\qquad \rho^{-}K^{0},
\]
and corresponding vector-meson channels. Due to their exotic isospin value $I=\frac{3}{2}$, these states cannot mix strongly with ordinary kaonic resonances and may therefore appear as distinct enhancements in the $\pi K$ and $\rho K$ invariant mass distributions.

The isotensor states with hypercharge $Y=0$ and isospin $I=2$,
\[
uu\bar{d}\bar{d}, \qquad uu\bar{u}\bar{d}, \qquad ud\bar{u}\bar{d}, \qquad dd\bar{u}\bar{d}, \qquad dd\bar{u}\bar{u},
\]
are particularly important since conventional $q\bar q$ mesons cannot possess isospin $I=2$. Consequently, these configurations represent unambiguous exotic candidates. Their production may occur in pion-induced reactions and central hadronic collisions involving large isospin transfer. Processes such as
\[
\pi^{+}p \rightarrow X^{++} + n,
\]
or double-charge exchange reactions may significantly enhance the production probability of isotensor tetraquarks.

The dominant strong decays of these states are expected through
\[
\pi\pi,\qquad \rho\rho,\qquad \pi\rho,
\]
channels. The doubly charged configuration $uu\bar{d}\bar{d}$ may decay into
\[
\pi^{+}\pi^{+},\qquad \rho^{+}\rho^{+},
\]
which provide exceptionally clean exotic signatures because no ordinary meson resonance can decay into such final states with identical charges. Likewise, the neutral isotensor states may decay into
\[
\pi^{+}\pi^{-},\qquad \pi^{0}\pi^{0},\qquad \rho^{+}\rho^{-},
\]
depending upon the corresponding isospin coupling. The large phase space available for these states suggests that their decay widths may be sizable, particularly for vector-meson channels.

The hidden-strangeness configurations with $Y=0$ and $I=1$,
\[
us\bar{u}\bar{s},\qquad
\frac{us\bar{d}\bar{s}+ds\bar{u}\bar{s}}{\sqrt{2}},\qquad
ds\bar{d}\bar{s},
\]
together with the mixed flavor combinations
\[
\frac{uu\bar{u}\bar{u}-ss\bar{s}\bar{s}}{\sqrt{2}},\qquad
\frac{ud\bar{u}\bar{d}-us\bar{u}\bar{s}}{\sqrt{2}},\qquad
\frac{dd\bar{d}\bar{d}-ds\bar{d}\bar{s}}{\sqrt{2}},
\]
may be produced abundantly in gluon-rich processes such as radiative $J/\psi$ decays,
\[
J/\psi \rightarrow \gamma X,
\]
where hidden-strangeness multiquark components are expected to be enhanced. Since these states contain both light and strange quarks, they may strongly couple to strange meson pairs.

Their dominant strong decay channels are expected to be
\[
K\bar K,\qquad K\bar K^{*},\qquad K^{*}\bar K^{*},\qquad \eta\pi,\qquad \eta'\pi.
\]
For example, the state $us\bar{u}\bar{s}$ can decay into
\[
K^{+}K^{-},\qquad K^{*+}K^{-},\qquad \phi\rho,
\]
while the mixed configuration
\[
\frac{us\bar{d}\bar{s}+ds\bar{u}\bar{s}}{\sqrt{2}},
\]
may couple to
\[
K^{+}\bar K^{0},\qquad K^{*+}\bar K^{0},
\]
and related strange-vector channels. The presence of hidden strangeness may also enhance couplings to $\eta$- and $\phi$-meson final states.

The flavor-singlet dominated states,
\[
\frac{uu\bar{u}\bar{u}+dd\bar{d}\bar{d}+4ss\bar{s}\bar{s}-2uu\bar{d}\bar{d}-2dd\bar{u}\bar{u}}{\sqrt{24}},
\]
and
\[
\frac{uu\bar{u}\bar{u}+dd\bar{d}\bar{d}-2ss\bar{s}\bar{s}}{\sqrt{6}},
\]
with $(I,Y)=(0,0)$, may strongly couple to scalar-isoscalar mesonic channels. Their dominant strong decay modes are expected through
\[
\eta\eta,\qquad \eta\eta',\qquad \phi\phi,\qquad K\bar K,
\]
and possibly through multi-pion final states. Because these states possess significant flavor mixing, they may exhibit broader decay widths and stronger overlap with conventional scalar and tensor mesons.

Finally, the negative hypercharge states with $Y=-1$ and $I=\frac{3}{2}$,
\[
su\bar{u}\bar{u},\qquad
su\bar{u}\bar{d},\qquad
sd\bar{u}\bar{d},\qquad
sd\bar{d}\bar{d},
\]
are expected to be produced in strange hadronic interactions and kaon-beam experiments. Their dominant decay channels include
\[
\bar K\pi,\qquad \bar K\rho,\qquad \bar K^{*}\pi,\qquad \bar K^{*}\rho.
\]
Similarly, the doubly strange states with $Y=-2$,
\[
ss\bar{u}\bar{u},\qquad
\frac{ss(\bar{u}\bar{d}+\bar{d}\bar{u})}{\sqrt{2}},\qquad
ss\bar{d}\bar{d},
\]
are expected to decay predominantly into
\[
\bar K\bar K,\qquad \bar K\bar K^{*},\qquad \bar K^{*}\bar K^{*},
\]
final states. The observation of such doubly strange exotic structures would provide strong evidence for the existence of compact multiquark dynamics beyond the conventional quark model.

\section{Summary and Conclusions}
In the present work, a systematic spectroscopic investigation of fully light tetraquark states belonging to the SU(3) flavor $\mathbf{27}$-plet has been performed within the framework of the extended Gursey-Radicati mass formula. The study has been focused on fully light exotic configurations of the type $(qq\bar q \bar q)$ possessing symmetric spin structure with spin-parity assignment $J^{P}=2^{+}$. The complete flavor multiplet structure has been constructed using SU(3) flavor symmetry and Young tableau techniques, allowing the classification of all possible tetraquark configurations according to their corresponding quantum numbers.

The spin wavefunctions of the tetraquark system were formulated through the Young tableau formalism by considering the permutation symmetry of four constituent particles. The fully symmetric spin configuration corresponding to total spin $S=2$ was identified and employed for the construction of the $\mathbf{27}$-plet states. Using the extended Gürsey--Radicati mass relation, the masses of all tetraquark members were calculated by incorporating the contributions from isospin, hypercharge, and the quadratic Casimir operator of SU(3) flavor symmetry. Since all states belong to the same SU(3) multiplet, the value of the Casimir operator remains fixed at
\[
C_{2}(SU(3)) = 8,
\]
while the mass splittings arise primarily due to flavor symmetry breaking effects associated with strange quark contributions.

The obtained mass spectra exhibit a clear hierarchical pattern governed by hypercharge and strangeness content. The lightest states were found in the $Y=+2$ sector with masses around $1.84~\mathrm{GeV}$, whereas the heaviest states correspond to the doubly strange configurations in the $Y=-2$ sector with masses near $2.47~\mathrm{GeV}$. The isotensor states with $(I,Y)=(2,0)$ were predicted around $2.32~\mathrm{GeV}$ and constitute particularly important exotic candidates because such quantum numbers cannot be realized within the conventional quark--antiquark meson picture. The results further indicate the presence of partial degeneracies among several neutral hidden-strangeness configurations, suggesting nontrivial flavor mixing effects within the tetraquark system.

In addition to the spectroscopic analysis, the possible production mechanisms and strong decay patterns of the $\mathbf{27}$-plet tetraquarks were investigated. The present analysis suggests that these exotic states may be produced in gluon-rich hadronic environments such as proton--proton collisions, heavy-ion collisions, radiative quarkonium decays, and kaon-induced reactions. Owing to their multiquark nature, the dominant strong decays are expected to proceed through OZI-superallowed fall-apart mechanisms into two-meson final states. Depending upon their flavor compositions and quantum numbers, the tetraquark states predominantly decay into channels involving
\[
KK,\quad K\bar K,\quad \pi K,\quad \pi\pi,\quad \rho\rho,\quad K^{*}K^{*},
\]
and other strange and nonstrange vector-meson combinations. In particular, the isotensor states and doubly strange configurations provide exceptionally clean experimental signatures due to their manifestly exotic flavor structures and suppressed mixing with ordinary mesons.

Overall, the present study provides a comprehensive theoretical framework for understanding fully light tetraquark states within the SU(3) flavor $\mathbf{27}$-plet. The predicted mass spectra, exotic quantum number assignments, and strong decay patterns may serve as useful theoretical guidance for future experimental investigations at facilities such as BESIII, COMPASS, GlueX, PANDA, and the LHC experiments. The observation of the predicted exotic states, especially those carrying isotensor or doubly strange quantum numbers, would provide strong evidence for the existence of compact multiquark configurations beyond the conventional quark model and would contribute significantly toward understanding the nonperturbative dynamics of Quantum Chromodynamics in the low-energy regime.

\nocite{*}

\bibliography{light}
\bibliographystyle{unsrt}

\end{document}